\documentclass[a4paper,12pt]{article}

\usepackage{amsmath}
\usepackage{amsthm}
\usepackage{amssymb}
\newtheorem{thm}{Theorem}
\newtheorem{prop}[thm]{Proposition}
\newtheorem{lem}[thm]{Lemma}

\theoremstyle{definition}
\newtheorem{defn}[thm]{Definition}

\theoremstyle{remark}

  \newtheorem{df-thm}[thm]{Definition-Theorem}
  \newtheorem{df-prop}[thm]{Definition-Proposition}
  \newtheorem{df-lem}[thm]{Definition-Lemma}
  \newtheorem{df-exa}[thm]{Definition-Example}

  \makeatletter
   
   \@addtoreset{equation}{section}
  \makeatother
\title{Some Formulas for Invariant Phases of \\
Unitary Matrices by Jarlskog}
\author{
Tatsuo SUZUKI{\thanks{E-mail address: 
i027110@sic.shibaura-it.ac.jp, 
 suzukita@aoni.waseda.jp}}\\
\\
Center for Educational Assistance\\
Shibaura Institute of Technology\\
Saitama, 337-8570\\
Japan}
\date{}
\begin{document}
\maketitle

\begin{abstract}
We describe calculations of Jarlskog's determinant in the case of 
$n=3,4$ in detail. Next, we investigate some formulas for invariant 
phases of unitary matrices and derive some explicit relations of them. 
\end{abstract}

\section{Introduction}
CP violation is expected in the standard model of particle physics 
with three or more families \cite{C}, \cite{KM}. Therefore it is an 
important problem that 
what is the measure of CP violation with such families 
which is invariant under the action of phase factors. 

To construct invariants for matrix action, the determinant is a useful tool. 
In the previous paper \cite{J1}, 
C. Jarlskog succeeded to define invariants of CP violation by using 
the determinant for commutator of the quark mass matrices. 
For the case of 3 families, it is relatively easy to calculate it. 
Moreover, in that case, her determinant is propotional to 
an invariant phase of unitary matrices. 
Then she discussed invariant quantities 
for $4$ families by using projection operators and the trace of 
some matrices, but she did not deal with her determinant 
itself for $n=4$ \cite{J2}. Therefore the problem is still remained. 
An approach to this problem is to use a parametrization for unitary matrices. 
See \cite{J3-0}, \cite{J3}, \cite{F1}, \cite{FFK} and their references within. 
Geometric constructions of them are also studied, see, for example, 
\cite{JS}, \cite{GG}, \cite{F2}. 

In this paper, we show the explicit calculation of Jarlskog's determinant 
in the case of 4 families and we find that it is hard to use 
the determinant for investigations of CP violation in the case of $n=4$.
Next, we study Jarlskog's invariant phases of unitary matrices 
instead of the determinant. 
we give some useful formulas for them 
and derive the detailed dependency of them which was simply described 
in the previous paper \cite{J3}.

\section{Jarlskog's Determinant}

Let $H,H'$ be $n \times n$-matrices(``mass matrices"),
their eigenvalues(``masses of quarks") $a_{i}, \ b_{i} \ $ 
(all multiplicities are $1$), and their diagonalizations 
$\ H=UDU^{\dagger}, H'=U'D'U'^{\dagger} $(where $U,U'$ are unitary matrices). 
Then, we define Jarlskog's determinant $\det [H,H'] \ $as follows
\cite{J1}; 
we put a ``quark mixing matrix" $V=U^{\dagger}U'$, then we have 
\begin{eqnarray}
\det [H,H']&=&\det (HH'-H'H)\nonumber\\
&=&\det (UDU^{\dagger}U'D'U'^{\dagger}-U'D'U'^{\dagger}UDU^{\dagger})
\nonumber\\
&=&\det (DU^{\dagger}U'D'U'^{\dagger}U-U^{\dagger}U'D'U'^{\dagger}UD)
\nonumber\\
&=&\det (DVD'V^{\dagger}-VD'V^{\dagger}D). \label{eqn:JD} 
\end{eqnarray}

We denote $V_{ij}$ as the components of the unitary matrix $V$, then 
by a straightforward calculation, we have 
$$(DVD'V^{\dagger}-VD'V^{\dagger}D)_{ij}
=(a_i-a_j)\sum_k b_k V_{ik}\bar{V}_{jk}. 
$$

\noindent
{\textbf{(I)}} \ First we consider her determinant (\ref{eqn:JD}) 
in the case of $n=3$. This is a result of Jarlskog \cite{J1}.

\begin{prop}[Jarlskog]
\begin{equation}
\det [H,H']=2i \ TB\ \mbox{Im}  \ 
(V_{11}V_{22}\bar{V}_{12}\bar{V}_{21})
\end{equation}
where 
$$
T=\prod_{(i,j)=(1,2),(2,3),(3,1)}(a_i-a_j), \quad 
B=\prod_{(i,j)=(1,2),(2,3),(3,1)}(b_i-b_j).
$$
\end{prop}

\noindent
ProofFFirst, we have
\begin{eqnarray*}
&&\det (DVD'V^{\dagger}-VD'V^{\dagger}D)\\
&=&\prod_{(i,j)=(1,2),(2,3),(3,1)}
(a_i-a_j)\sum_{k=1}^3 b_k V_{ik}\bar{V}_{jk}
-\prod_{(i,j)=(1,2),(2,3),(3,1)}
(a_i-a_j)\sum_{k=1}^3 b_k V_{jk}\bar{V}_{ik}. 
\end{eqnarray*}

\noindent
Therefore, we obtain 
\begin{eqnarray*}
&&\det (DVD'V^{\dagger}-VD'V^{\dagger}D)/T\\
&=&\prod_{(i,j)=(1,2),(2,3),(3,1)}
\sum_{k=1}^3 b_k (V_{ik}\bar{V}_{jk}
- V_{jk}\bar{V}_{ik})\\
&=&2i \ \mbox{Im}  \prod_{(i,j)=(1,2),(2,3),(3,1)}
\sum_{k=1}^3 b_k V_{ik}\bar{V}_{jk}\\
&=&2i \ \mbox{Im}  \prod_{(i,j)=(1,2),(2,3),(3,1)}
\left(
\sum_{k=1}^2 b_k V_{ik}\bar{V}_{jk}+b_3 V_{i3}\bar{V}_{j3}
\right)\\
&=&2i \ \mbox{Im}  \prod_{(i,j)=(1,2),(2,3),(3,1)}
\left(
\sum_{k=1}^2 b_k V_{ik}\bar{V}_{jk}+b_3 
(\delta_{ij}-V_{i1}\bar{V}_{j1}-V_{i2}\bar{V}_{j3})
\right)\\
&=&2i \ \mbox{Im}  \prod_{(i,j)=(1,2),(2,3),(3,1)}
\left(
\sum_{k=1}^2 (b_k-b_3) V_{ik}\bar{V}_{jk}
\right) .
\end{eqnarray*}
%
%
\begin{eqnarray*}
&=&2i \ \mbox{Im}  
\left\{
\sum_{k_1=1}^2 (b_{k_1}-b_3) V_{1k_1}\bar{V}_{2k_1}
\right\}
\left\{
\sum_{k_2=1}^2 (b_{k_2}-b_3) V_{2k_2}\bar{V}_{3k_2}
\right\}
\left\{
\sum_{k_3=1}^2 (b_{k_3}-b_3) V_{3k_3}\bar{V}_{1k_3}
\right\} \\
&=&2i \ \mbox{Im}  
\sum_{k_1,k_2,k_3=1}^2 
(b_{k_1}-b_3)(b_{k_2}-b_3)(b_{k_3}-b_3)
V_{1k_1}\bar{V}_{2k_1}
V_{2k_2}\bar{V}_{3k_2}
V_{3k_3}\bar{V}_{1k_3}
\\
&=&2i \ \mbox{Im}  
\sum_{k_1,k_2,k_3=1}^2 
(b_{k_1}-b_3)(b_{k_2}-b_3)(b_{k_3}-b_3)\\
&& \hspace{2cm} \times 
\left\{
V_{1k_1}\bar{V}_{2k_1}
V_{2k_2}\bar{V}_{1k_3}
(\delta_{k_2 k_3}-\bar{V}_{1k_2}{x}_{1k_3}-\bar{V}_{2k_2}{x}_{2k_3})
\right\} \\
&=&2i \ \mbox{Im}  
\sum_{k_1,k_2=1}^2 
(b_{k_1}-b_3)(b_{k_2}-b_3)^2 
V_{1k_1}\bar{V}_{2k_1}
V_{2k_2}\bar{V}_{1k_2}
\\
&&-2i \ \mbox{Im}  
\sum_{k_1,k_2,k_3=1}^2 
(b_{k_1}-b_3)(b_{k_2}-b_3)(b_{k_3}-b_3)\\
&& \hspace{2cm} \times 
\left\{
V_{1k_1}\bar{V}_{2k_1}
V_{2k_2}\bar{V}_{1k_2}
|{x}_{1k_3}|^2
+V_{1k_1}\bar{V}_{2k_1}
{x}_{2k_3}\bar{V}_{1k_3}
|V_{2k_2}|^2)
\right\} .
\end{eqnarray*}

\newpage

\noindent
We calculate all sums explicitly and we remark 
that 
$V_{11}\bar{V}_{21}
V_{22}\bar{V}_{12}
+V_{12}\bar{V}_{22}
V_{21}\bar{V}_{11}$ is real and its imaginary part vanishes, then 
we have 
\begin{eqnarray*}
&&2i \ \mbox{Im}  \ 
\left\{ 
(b_{1}-b_3)(b_{2}-b_3)^2 
V_{11}\bar{V}_{21}
V_{22}\bar{V}_{12}
+
(b_{2}-b_3)(b_{1}-b_3)^2 
V_{12}\bar{V}_{22}
V_{21}\bar{V}_{11}
\right\} \\
&=&
2i \ \mbox{Im}  \ 
\left\{ 
(b_{1}-b_3)(b_{2}-b_3)^2 
V_{11}\bar{V}_{21}
V_{22}\bar{V}_{12}
-
(b_{2}-b_3)(b_{1}-b_3)^2 
V_{11}\bar{V}_{21}
V_{22}\bar{V}_{12}
\right\} \\
&=&
2i \ \mbox{Im}  \ 
\left\{ 
(b_{1}-b_3)(b_{2}-b_3)^2 
-
(b_{2}-b_3)(b_{1}-b_3)^2 
\right\} 
V_{11}\bar{V}_{21}
V_{22}\bar{V}_{12}
\\
&=&
2i \ \mbox{Im}  \ 
(b_{1}-b_3)(b_{2}-b_3)(b_2-b_1)
V_{11}\bar{V}_{21}
V_{22}\bar{V}_{12}
\end{eqnarray*}
Therfore we conclude the proof. 

\bigskip 

\noindent
{\textbf{(II)} \ Next, we consider (\ref{eqn:JD}) in the case of $n=4$. 
As mensioned above, Jarlskog did not deal with her determinant 
(\ref{eqn:JD}) itself for $n=4$. 
Therefore we calculate it directly. 

Here we put 
$$u_{ij}:= 
(DVD'V^{\dagger}-VD'V^{\dagger}D)_{ij}
=(a_i-a_j)\sum_k b_k V_{ik}\bar{V}_{jk}, 
\quad (\mbox{where} \  u_{ji}=-\bar{u}_{ij}) .$$
Then we have 
\begin{eqnarray*}
&&\det (DVD'V^{\dagger}-VD'V^{\dagger}D)
=
\left|
\begin{array}{cccc}
0 & u_{12} & u_{13} & u_{14} \\
-\bar{u}_{12} & 0 & u_{23} & u_{24} \\
-\bar{u}_{13} & -\bar{u}_{23} & 0 & u_{34} \\
-\bar{u}_{14} & -\bar{u}_{24} & -\bar{u}_{34} & 0\\
\end{array}
\right| \\
&& \\
&=&{u}_{12}{u}_{34}\bar{u}_{12}\bar{u}_{34}
+{u}_{13}{u}_{24}\bar{u}_{13}\bar{u}_{24}
+{u}_{14}{u}_{23}\bar{u}_{14}\bar{u}_{23}\\
&& \hspace{1cm} 
-\{ 
{u}_{12}{u}_{24}\bar{u}_{13}\bar{u}_{34}
+{u}_{13}{u}_{34}\bar{u}_{12}\bar{u}_{24}
+
{u}_{14}{u}_{23}\bar{u}_{13}\bar{u}_{24}
+{u}_{13}{u}_{24}\bar{u}_{14}\bar{u}_{23} \} 
\\
&& \hspace{1cm} 
+{u}_{14}\bar{u}_{12}\bar{u}_{23}\bar{u}_{34}
+{u}_{12}{u}_{23}{u}_{34}\bar{u}_{14}
\\
\end{eqnarray*}
\begin{eqnarray*}
&=&\left\{ \left(
\prod_{(i,j)=(1,2),(3,4),(2,1),(4,3)}
+\prod_{(i,j)=(1,3),(2,4),(3,1),(4,2)}
+\prod_{(i,j)=(1,4),(2,3),(4,1),(3,2)} 
\right) \right. \\
&& \hspace{-1cm} 
-\left(
\prod_{(i,j)=(1,2),(2,4),(3,1),(4,3)}
+\prod_{(i,j)=(1,3),(3,4),(2,1),(4,2)}
+\prod_{(i,j)=(1,4),(2,3),(3,1),(4,2)} 
+\prod_{(i,j)=(1,3),(2,4),(4,1),(3,2)} 
\right) \\
&& \hspace{1cm} 
-\left. \left(
\prod_{(i,j)=(1,4),(2,1),(3,2),(4,3)}
+\prod_{(i,j)=(1,2),(2,3),(3,4),(4,1)}
\right) \right\} 
(a_i-a_j)\sum_{k=1}^4 b_k V_{ik}\bar{V}_{jk}
\\
&& \\
&=&\mbox{Re}\left[ \left\{ \left(
\prod_{(i,j)=(1,2),(3,4),(2,1),(4,3)}
+\prod_{(i,j)=(1,3),(2,4),(3,1),(4,2)}
+\prod_{(i,j)=(1,4),(2,3),(4,1),(3,2)} 
\right) \right. \right. \\
&& \hspace{-1cm} -2 \left.
\left(
\prod_{(i,j)=(1,2),(2,4),(3,1),(4,3)}
+\prod_{(i,j)=(1,3),(2,4),(4,1),(3,2)} 
+\prod_{(i,j)=(1,2),(2,3),(3,4),(4,1)}
\right) \right\}
\\
&& \hspace{3cm} 
\left. 
(a_i-a_j)\sum_{k=1}^4 b_k V_{ik}\bar{V}_{jk}
\right] .
\end{eqnarray*}
Here we put 
$$T_{(ij)(kl)}:=(a_i-a_j)^2(a_k-a_l)^2, \quad 
T_{(ijkl)}:=(a_i-a_j)(a_j-a_k)(a_k-a_l)(a_l-a_i)$$
$$B_{(ij)(kl)}:=(b_i-b_j)^2(b_k-b_l)^2, \quad 
B_{(ijkl)}:=(b_i-b_j)(b_j-b_k)(b_k-b_l)(b_l-b_i)$$
$$b_{k4}:=b_k-b_4$$
and we remark 
$$V_{i4}\bar{V}_{j4}
=\delta_{ij}-\sum_{k=1}^3 V_{ik}\bar{V}_{jk}
=-\sum_{k=1}^3 V_{ik}\bar{V}_{jk} \quad (i \ne j), $$
then, 
\begin{eqnarray*}
&&\hspace{-25mm}
=\mbox{Re} \left[ \left(
T_{(12)(34)}\prod_{(i,j)=(1,2),(3,4),(2,1),(4,3)}
+T_{(13)(24)}\prod_{(i,j)=(1,3),(2,4),(3,1),(4,2)}
+T_{(14)(23)}\prod_{(i,j)=(1,4),(2,3),(4,1),(3,2)} 
\right) \right. \\
&& \hspace{-15mm} -2 
\left\{ 
\left(
T_{(1243)}\prod_{(i,j)=(1,2),(2,4),(3,1),(4,3)}
+T_{(1324)}\prod_{(i,j)=(1,3),(2,4),(4,1),(3,2)} 
+T_{(1234)}\prod_{(i,j)=(1,2),(2,3),(3,4),(4,1)}
\right)
\right\} \\
&& \hspace{3cm} 
\left. 
\sum_{k=1}^3 b_{4k} V_{ik}\bar{V}_{jk}
 \right]
\\
&&\hspace{-25mm}
=\mbox{Re} \left[ 
T_{(12)(34)}\sum_{k_1,\cdots,k_4=1}^3
b_{k_1 4}\cdots b_{k_4 4} \ \ 
V_{1k_1}\bar{V}_{2k_1}V_{3k_2}\bar{V}_{4k_2}
V_{2k_3}\bar{V}_{1k_3}V_{4k_4}\bar{V}_{3k_4}
\right. \\
&& \hspace{-15mm}
+T_{(13)(24)}\sum_{k_1,\cdots,k_4=1}^3
b_{k_1 4}\cdots b_{k_4 4} \ \ 
V_{1k_1}\bar{V}_{3k_1}V_{2k_2}\bar{V}_{4k_2}
V_{3k_3}\bar{V}_{1k_3}V_{4k_4}\bar{V}_{2k_4}
\\
&& \hspace{-15mm}
+T_{(14)(23)}\sum_{k_1,\cdots,k_4=1}^3
b_{k_1 4}\cdots b_{k_4 4} \ \ 
V_{1k_1}\bar{V}_{4k_1}V_{2k_2}\bar{V}_{3k_2}
V_{4k_3}\bar{V}_{1k_3}V_{3k_4}\bar{V}_{2k_4}
\\
&& \hspace{-15mm} -2 
\left(
T_{(1243)}\sum_{k_1,\cdots,k_4=1}^3
b_{k_1 4}\cdots b_{k_4 4} \ \ 
V_{1k_1}\bar{V}_{2k_1}V_{2k_2}\bar{V}_{4k_2}
V_{3k_3}\bar{V}_{1k_3}V_{4k_4}\bar{V}_{3k_4}
\right.
\\
&& \hspace{-8mm} 
+T_{(1324)}\sum_{k_1,\cdots,k_4=1}^3
b_{k_1 4}\cdots b_{k_4 4} \ \ 
V_{1k_1}\bar{V}_{3k_1}V_{2k_2}\bar{V}_{4k_2}
V_{4k_3}\bar{V}_{1k_3}V_{3k_4}\bar{V}_{2k_4}
\\
&& \hspace{-8mm} \left. \left. 
+T_{(1234)}\sum_{k_1,\cdots,k_4=1}^3
b_{k_1 4}\cdots b_{k_4 4} \ \ 
V_{1k_1}\bar{V}_{2k_1}V_{2k_2}\bar{V}_{3k_2}
V_{3k_3}\bar{V}_{4k_3}V_{4k_4}\bar{V}_{1k_4}
\right) \right]
\end{eqnarray*}
Here we introduce some notations; 
$$[\alpha \beta ; jk]:=
V_{\alpha j}V_{\beta k}\bar{V}_{\alpha k}\bar{V}_{\beta j}, 
\qquad 
(abc)=(abc;k_1 k_2 k_3):=V_{ak_1}\bar{V}_{bk_1}
V_{bk_2}\bar{V}_{ck_2}V_{ck_3}\bar{V}_{ak_3}. 
$$
Moreover by using a relation 
$$ V_{4k_2}\bar{V}_{4k_4}
=\delta_{k_2 k_4}-V_{1k_2}\bar{V}_{1k_4}
-V_{2k_2}\bar{V}_{2k_4}-V_{3k_2}\bar{V}_{3k_4}, 
$$
then, for example, the coefficient of $\ T_{(12)(34)}\ $ is 
\begin{eqnarray*}
&&\hspace{-15mm} 
\sum_{k_1,\cdots,k_4=1}^3
b_{k_1 4}\cdots b_{k_4 4} \ \ 
V_{1k_1}\bar{V}_{2k_1}V_{3k_2}\bar{V}_{4k_2}
V_{2k_3}\bar{V}_{1k_3}V_{4k_4}\bar{V}_{3k_4}
\\
&&\hspace{-15mm} 
=\sum_{k_1,\cdots,k_4=1}^3
b_{k_1 4}\cdots b_{k_4 4} \ 
[12;k_1 k_3] \ [34;k_2 k_4]
\\
&&\hspace{-15mm} 
=\sum_{k_1,k_2,k_3=1}^3
b_{k_1 4}b_{k_2 4}^2 b_{k_3 4} \ 
[12;k_1 k_3] |V_{3 k_2}|^2 
-\sum_{k_1,\cdots,k_4=1}^3
b_{k_1 4}\cdots b_{k_4 4} \ 
[12;k_1 k_3] \ [31;k_2 k_4] 
\\
&& -\sum_{k_1,\cdots,k_4=1}^3
b_{k_1 4}\cdots b_{k_4 4} \ 
[12;k_1 k_3] \ [32;k_2 k_4]
-\sum_{k_1,\cdots,k_4=1}^3
b_{k_1 4}\cdots b_{k_4 4} \ 
[12;k_1 k_3] \ [33;k_2 k_4]
\\
&&\hspace{-15mm} 
=\sum_{k_1,k_2,k_3=1}^3
b_{k_1 4}b_{k_2 4}b_{k_3 4}^2 \ 
[12;k_1 k_2] |V_{3 k_3}|^2 
-\sum_{k_1,\cdots,k_4=1}^3
b_{k_1 4}\cdots b_{k_4 4} \ 
[12;k_1 k_2] \ [13;k_3 k_4] 
\\
&& -\sum_{k_1,\cdots,k_4=1}^3
b_{k_1 4}\cdots b_{k_4 4} \ 
[12;k_1 k_2] \ [23;k_3 k_4]
-\sum_{k_1,\cdots,k_4=1}^3
b_{k_1 4}\cdots b_{k_4 4} \ 
[12;k_1 k_3] \ |V_{3 k_2}|^2 |V_{3 k_4}|^2
\\
&&\hspace{-20mm}
\\
\end{eqnarray*}
Furthermore, the coefficient of $\ -2 T_{(1243)}\ $ is 
\begin{eqnarray*}
&&\hspace{-15mm} 
\sum_{k_1,\cdots,k_4=1}^3
b_{k_1 4}\cdots b_{k_4 4} \ \ 
V_{1k_1}\bar{V}_{2k_1}V_{2k_2}\bar{V}_{4k_2}
V_{3k_3}\bar{V}_{1k_3}V_{4k_4}\bar{V}_{3k_4}
\\
&&\hspace{-15mm} 
=\sum_{k_1,k_2,k_3=1}^3
b_{k_1 4}b_{k_2 4}^2 b_{k_3 4} \ 
(123;k_1 k_2 k_3)
-\sum_{k_1,\cdots,k_4=1}^3
b_{k_1 4}\cdots b_{k_4 4} \ 
[12;k_1 k_2] \ [31;k_3 k_4] 
\\
&& 
-\sum_{k_1,\cdots,k_4=1}^3
b_{k_1 4}\cdots b_{k_4 4} \ 
((123;k_1 k_4 k_3)|V_{2k_2}|^2
+(123;k_1 k_2 k_3)|V_{3k_4}|^2)
\\
&&\hspace{-15mm} 
=\sum_{k_1,k_2,k_3=1}^3
b_{k_1 4}b_{k_2 4} b_{k_3 4}^2 \ 
(312)
-\sum_{k_1,\cdots,k_4=1}^3
b_{k_1 4}\cdots b_{k_4 4} \ 
[12;k_1 k_2] \ [13;k_3 k_4] 
\\
&& 
-\sum_{k_1,\cdots,k_4=1}^3
b_{k_1 4}\cdots b_{k_4 4} \ 
(123)(|V_{2k_4}|^2+|V_{3k_4}|^2). 
\end{eqnarray*}
Then, we can sum up the coefficient of 
$
-\sum_{k_1,\cdots,k_4=1}^3
b_{k_1 4}\cdots b_{k_4 4} \ 
[12;k_1 k_2] \ [13;k_3 k_4]$; 
$$
T_{(12)(34)}+T_{(13)(24)}-2T_{(1243)}=T_{(14)(23)}, 
$$
similary, the coefficient of $
-\sum_{k_1,\cdots,k_4=1}^3
b_{k_1 4}\cdots b_{k_4 4} \ 
[12;k_1 k_2] \ [23;k_3 k_4] $; 
$$
T_{(12)(34)}+T_{(14)(23)}-2T_{(1234)}=T_{(13)(24)}, 
$$
and the coefficient of 
$
-\sum_{k_1,\cdots,k_4=1}^3
b_{k_1 4}\cdots b_{k_4 4} \ 
[13;k_1 k_2] \ [23;k_3 k_4] $, 
$$
T_{(13)(24)}+T_{(14)(23)}-2T_{(1324)}=T_{(12)(34)}$$
Therefore we obtain the following theorem; 

\newpage
\begin{thm}
\begin{eqnarray*}
&&\hspace{-20mm}
\det (DVD'V^{\dagger}-VD'V^{\dagger}D)\\
&&\hspace{-25mm}
=\mbox{Re}\left[ 
T_{(12)(34)}
\sum_{k_1,k_2,k_3=1}^3
b_{k_1 4}b_{k_2 4}b_{k_3 4}^2 \ 
[12;k_1 k_2] |V_{3 k_3}|^2 
\right.
\\
&&\hspace{-15mm}
+T_{(13)(24)}
\sum_{k_1,k_2,k_3=1}^3
b_{k_1 4}b_{k_2 4}b_{k_3 4}^2 \ 
[13;k_1 k_2] |V_{2 k_3}|^2 
\\
&&\hspace{-15mm}
+T_{(14)(23)}
\sum_{k_1,k_2,k_3=1}^3
b_{k_1 4}b_{k_2 4}b_{k_3 4}^2 \ 
[23;k_1 k_2] |V_{1 k_3}|^2 
\\
&&\hspace{-25mm}
-2\left( 
T_{(1243)}
\sum_{k_1,k_2,k_3=1}^3
b_{k_1 4}b_{k_2 4} b_{k_3 4}^2 \ 
(312)
+T_{(1324)}
\sum
b_{k_1 4}b_{k_2 4} b_{k_3 4}^2 \ 
(132)
+T_{(1234)}
\sum
b_{k_1 4}b_{k_2 4} b_{k_3 4}^2 \ 
(123)
\right)
\\
&&\hspace{-15mm}
-\left\{ 
T_{(12)(34)}\sum_{k_1,\cdots,k_4=1}^3
b_{k_1 4}\cdots b_{k_4 4} \ 
\left( 
[13;k_1 k_2] \ [23;k_3 k_4] 
+[12;k_1 k_2] \ |V_{3 k_3}|^2 |V_{3 k_4}|^2
\right)
\right.
\\
&&\hspace{-9mm}
+T_{(13)(24)}
\sum_{k_1,\cdots,k_4=1}^3
b_{k_1 4}\cdots b_{k_4 4} \ 
\left( 
[12;k_1 k_2] \ [23;k_3 k_4] 
+[13;k_1 k_2] \ |V_{2 k_3}|^2 |V_{2 k_4}|^2
\right)
\\
&&\hspace{-9mm}
+T_{(14)(23)}
\sum_{k_1,\cdots,k_4=1}^3
b_{k_1 4}\cdots b_{k_4 4} \ 
\left( 
[12;k_1 k_2] \ [13;k_3 k_4] 
+[23;k_1 k_2] \ |V_{1 k_3}|^2 |V_{1 k_4}|^2
\right)
\\
&&\hspace{-9mm}
-2\left( 
T_{(1243)}
\sum_{k_1,\cdots,k_4=1}^3
b_{k_1 4}\cdots b_{k_4 4} \ 
(312)(|V_{2k_4}|^2+|V_{3k_4}|^2)
\right. 
\\
&&\hspace{-1mm}
+T_{(1324)}
\sum_{k_1,\cdots,k_4}^3
b_{k_1 4}\cdots b_{k_4 4} \ 
(132)(|V_{1k_4}|^2+|V_{2k_4}|^2)
\\
&&\hspace{-2mm}
\left. \left. \left.
+T_{(1234)}
\sum_{k_1,\cdots,k_4}^3
b_{k_1 4}\cdots b_{k_4 4} \ 
(123)(|V_{1k_4}|^2+|V_{3k_4}|^2)
\right) \right\} \right]
\\
\end{eqnarray*}
\end{thm}
\noindent
\textbf{Remark}\ 
We have a relation 
$$
T_{(12)(34)}+T_{(13)(24)}+T_{(14)(23)}
-2T_{(1243)}-2T_{(1324)}-2T_{(1234)}=0. $$
However, we cannot sum up this determinant to more compact form 
any more. Therfore we conclude that, in case of $n=4$, it is hard to use 
Jarlskog's determinant for investigations of CP violation. 

\newpage

\section{Invariant Phases of Unitary Matrices}
To study CP violation, we need quantities which are invariant 
under the action
$$V \to diag(e^{i\theta_1}, \cdots, e^{i\theta_n}) \ V \ 
diag(e^{i\theta'_1}, \cdots, e^{i\theta'_n}) . $$
One of them is Jarlskog's determinant $\ \det [H,H']\ $ and 
in the case of $n=3$ it has a simple form 
$$
\det [H,H']=2i \ TB\ \mbox{Im}  \ 
(V_{11}V_{22}\bar{V}_{12}\bar{V}_{21}).
$$
However, as we showed in the previous section, 
the determinant is much complicated and 
it is hard to use it in the case of $n \geq 4$. 
Therefore, according to \cite{J1}, we introduce invariant phases 
of unitary matrices. 
\begin{defn}
$$(\alpha \beta ; jk):=
\mbox{Im} \ (V_{\alpha j}V_{\beta k}\bar{V}_{\alpha k}\bar{V}_{\beta j}),
$$
$$
<\alpha \beta ; jk>:=
\mbox{Re} \ (V_{\alpha j}V_{\beta k}\bar{V}_{\alpha k}\bar{V}_{\beta j}).
$$
\end{defn}
First we have 
\begin{lem}
$$(\alpha \beta ; kj)=-(\alpha \beta ; jk), \quad 
(\beta \alpha ; jk)=-(\alpha \beta ; jk)\quad 
(\mbox{antisymmetric w.r.t. }\alpha \ and \ \beta, \ j \ and \ k)
$$
$$<\alpha \beta ; kj>=<\alpha \beta ; jk>, \quad 
<\beta \alpha ; jk>=<\alpha \beta ; jk>\ 
(\mbox{symmetric w.r.t. }\alpha \ and \ \beta, \ j \ and \ k).
$$
\end{lem}

\noindent
Proof: The proof is easy. 

\bigskip

In case of $n=3$, we have already showed 
$\ \det [H,H']=2i \ TB\ (12;12)\ $. 
To investigate relations of $ (\alpha \beta ; jk) $s or 
$ <\alpha \beta ; jk> $s, 
the following proposition is fundamental. 
\begin{prop}
\ {\textbf{(I)}} \ (Unitary relations of imaginary part)
$$\sum_{\alpha=1}^n (\alpha \beta ; jk)=0, \quad 
\sum_{\beta=1}^n (\alpha \beta ; jk)=0 \quad 
\mbox{(row unitary relations)},$$
$$\sum_{j=1}^n (\alpha \beta ; jk)=0, \quad 
\sum_{k=1}^n (\alpha \beta ; jk)=0 \quad 
\mbox{(column unitary relations)}.$$
\noindent
{\textbf{(II)}} \ (Unitary relations of real part)
$$\sum_{\alpha=1}^n <\alpha \beta ; jk>=\delta_{jk}|V_{\beta j}|^2, 
\quad 
\sum_{\beta=1}^n <\alpha \beta ; jk>=\delta_{jk}|V_{\alpha j}|^2
\quad 
\mbox{(row unitary relations)},$$
$$\sum_{j=1}^n <\alpha \beta ; jk>=\delta_{\alpha \beta}|V_{\alpha k}|^2, 
\quad 
\sum_{k=1}^n <\alpha \beta ; jk>=\delta_{\alpha \beta}|V_{\alpha j}|^2
\quad 
\mbox{(column unitary relations)}.$$
\label{prop:unitary}
\end{prop}
\noindent
Proof: 
We only prove $\displaystyle \sum_{\alpha=1}^n (\alpha \beta ; jk)=0$. 
Noting that unitary relations 
$\displaystyle \sum_{\alpha=1}^n V_{\alpha j}\bar{V}_{\alpha k}
=\delta_{jk}$, we have
\begin{eqnarray*}
\sum_{\alpha=1}^n (\alpha \beta ; jk)
&=&\mbox{Im} \ 
\sum_{\alpha=1}^n (V_{\alpha j}V_{\beta k}\bar{V}_{\alpha k}\bar{V}_{\beta j})
\\
&=&\mbox{Im} \ 
(\sum_{\alpha=1}^n (V_{\alpha j}\bar{V}_{\alpha k})
V_{\beta k}\bar{V}_{\beta j})
\\
&=&\mbox{Im} \ 
(\delta_{jk}
V_{\beta k}\bar{V}_{\beta j})
\\
&=&0.
\end{eqnarray*}
We can prove other relations in a similar way. 
Therfore we conclude the proof. 

\bigskip

By using the proposition \ref{prop:unitary}, in case of $n=3$, 
if we remark relations 
$(\alpha \beta ; jj)=0$, then we have 
$$(12;13)=-(12;11)-(12;12)=-(12;12), \quad 
(12;23)=-(12;21)=(12;12), $$
$$(13;12)=-(12;12), \quad 
(13;13)=-(13;12)=(12;12), $$
$$(13;23)=-(13;21)=(12;21)=-(12;12), $$
$$(23;12)=-(21;12)=(12;12), \quad 
(23;13)=-(23;12)=(21;12)=-(12;12), $$
$$(23;23)=-(21;23)=(21;21)=(12;12) . $$
Therefore we have one independent invariant phase $\ (12;12) \ $. 

Next, in case of $n=4$, because of $(\alpha, \beta),(j,k)
=(1,2),(1,3),(1,4),(2,3),(2,4),(3,4)$, we have 
$6 \times 6=36$ possibilitiesDIn view of the theory for CP violation, 
we would like to find only three invariant phases in the case of $n=4$. 
To show this, according to \cite{J3}, we put 
$$R_{\alpha j}:=<\alpha, \alpha+1; j,j+1> \quad 
(\alpha, j=1,2,3),$$
$$J_{\alpha j}:=(\alpha, \alpha+1; j,j+1) \qquad 
(\alpha, j=1,2,3).$$
Then we have the following proposition. 

\newpage

\begin{prop}
We put 
$$\mathbf{J}=[J_{\alpha j}]_{
\tiny{
\left\{
\begin{array}{l}
\alpha=1,2,3\\
j=1,2,3\\
\end{array}
\right. }}
, \quad 
A=
\left[
\begin{array}{ccc}
1 & -1 & 0\\
-1 & 1 & -1\\
0 & -1 & 1\\
\end{array}
\right]
, $$
then 36 $(\alpha \beta ;jk)$s 
are expressed by 
$$\mathbf{J}, \ \mathbf{J}A, \ A\mathbf{J}, \ A\mathbf{J}A.$$
More explicitly, 
$$[(\alpha \beta ;jk)]_{
\tiny{
\left\{
\begin{array}{l}
(\alpha \beta)=(12),(23),(34)\\
(jk)=(24),(14),(13)\\
\end{array}
\right. }}
=\left[
\begin{array}{ccc}
(12;24) & (12;14) & (12;13)\\
(23;24) & (23;14) & (23;13)\\
(34;24) & (34;14) & (34;13)\\
\end{array}
\right]
=\mathbf{J}A,
$$
$$[(\alpha \beta ;jk)]_{
\tiny{
\left\{
\begin{array}{l}
(\alpha \beta)=(24),(14),(13)\\
(jk)=(12),(23),(34)\\
\end{array}
\right. }}
=A\mathbf{J}, 
\quad 
[(\alpha \beta ;jk)]_{
\tiny{
\left\{
\begin{array}{l}
(\alpha \beta)=(24),(14),(13)\\
(jk)=(24),(14),(13)\\
\end{array}
\right. }}
=A\mathbf{J}A.
$$
\label{prop:relation}
\end{prop}

\noindent
Proof: 
By using the proposition \ref{prop:unitary}, we have 
\begin{eqnarray}
0&=&(12;21)+(12;23)+(12;24) \nonumber\\
&=&-J_{11}+J_{12}+(12;24) \nonumber\\
(12;24)&=&J_{11}-J_{12}.
\label{eqn:1224}
\end{eqnarray}
\begin{eqnarray}
0&=&(12;31)+(12;32)+(12;34) \nonumber\\
&=&-(12;13)-J_{12}+J_{13} \nonumber\\
(12;13)&=&-J_{12}+J_{13}.
\end{eqnarray}
By using (\ref{eqn:1224}), 
\begin{eqnarray}
0&=&(12;41)+(12;42)+(12;43) \nonumber\\
&=&-(12;14)-(12;24)-J_{13} \nonumber\\
&=&-(12;14)-J_{11}+J_{12}-J_{13} \nonumber\\
(12;14)&=&-J_{11}+J_{12}-J_{13}.
\end{eqnarray}
In a similar way, we find that 36 $(\alpha \beta ;jk)$s 
are expressed by the linear combination 
of 9 $J_{\alpha j}$s. 
Therfore we conclude the proof. 

\bigskip 

By this proposition \ref{prop:relation}, 
we have only to show 9 $J_{\alpha j}$s as combinations of three of them, 
say $J_{11}, \ J_{22}, \ J_{33}\ $, 
we need more nonlinear relations between these invariant phases. 
The following proposition gives the relations of them. 
\begin{prop}
\ {\textbf{(I)}} \ 
\begin{equation}
<\alpha \beta ;jk>(\alpha \beta ;kl)
+<\alpha \beta ;kl>(\alpha \beta ;jk)
=<\alpha \beta ;kk>(\alpha \beta ;jl)
\label{eqn:row1}
\end{equation}
\begin{equation}
<\alpha \beta ;jk>(\beta \gamma ;jk)
+<\beta \gamma ;jk>(\alpha \beta ;jk)
=<\beta \beta ;jk>(\alpha \gamma ;jk)
\label{eqn:column1}
\end{equation}
\ {\textbf{(II)}} \ 
\begin{equation}
<\alpha \beta ;jk><\alpha \beta ;lm>
-<\alpha \beta ;jm><\alpha \beta ;kl>
=(\alpha \beta ;jl)(\alpha \beta ;km)
\label{eqn:row2}
\end{equation}
\begin{equation}
<\alpha \beta ;jk><\gamma \delta ;jk>
-<\alpha \delta ;jk><\beta \gamma ;jk>
=(\alpha \gamma ;jk)(\beta \delta ;jk)
\label{eqn:column2}
\end{equation}
\end{prop}
\noindent
Proof: We can prove them by a straightforward calculation. 

\bigskip

For example, we put $(\alpha \beta )=(12), 
j=1,k=2,l=3$ in (\ref{eqn:row1}), then 
\begin{eqnarray*}
<12;12>(12;23)+<12;23>(12;12)
&=&<12;22>(12;13)\\
R_{11}J_{12}+R_{12}J_{11}&=&|V_{12}V_{22}|^2(-J_{12}+J_{13})\\
-(|V_{12}V_{22}|^2+R_{11})J_{12}+|V_{12}V_{22}|^2 J_{13}
&=&R_{12}J_{11}
\end{eqnarray*}
Thus, we obtain the relations in \cite{J3} as follows;
$$
\hspace{-10mm}
{\tiny{
\left[
\begin{array}{cccccc}
-(|V_{12}V_{22}|^2+R_{11}) & |V_{12}V_{22}|^2 & 0 & 0 & 0 & 0 \\
0 & |V_{33}V_{34}|^2 & 0 & -(|V_{33}V_{34}|^2+R_{33}) & 0 & 0 \\
0 & 0 & -(|V_{21}V_{22}|^2+R_{11}) & 0 & |V_{21}V_{22}|^2 & 0 \\
0 & 0 & 0 & 0 & |V_{33}V_{43}|^2 & -(|V_{33}V_{43}|^2+R_{33}) \\
|V_{32}V_{33}|^2 & 0 & 0 & 0 & 0 & -R_{22} \\
0 & 0 & |V_{23}V_{33}|^2 & -R_{22} & 0 & 0 \\
\end{array}
\right]
\left[
\begin{array}{c}
J_{12} \\ J_{13} \\ J_{21} \\ J_{23} \\ J_{31} \\ J_{32}
\end{array}
\right] }}
$$
$$
=\left[
\begin{array}{c}
R_{12}J_{11} \\ R_{23}J_{33} \\ R_{21}J_{11} \\ R_{32}J_{33} \\ 
(|V_{32}V_{33}|^2+R_{32})J_{22} \\ (|V_{23}V_{33}|^2+R_{23})J_{22}
\end{array}
\right] .
$$
Since the rank of the coefficient matrix are generally six, 
we showed that 36 $(\alpha \beta ;jk)$s 
are expressed by $J_{11}, J_{22}$ and $J_{33}$.
\section{Discussion}
In this paper, we showed the explicit calculation of 
Jarlskog's determinant in the case of 4 families and 
we realized that it was hard to use 
the determinant for investigations of CP violation in the case of $n \geq 4$.
Next, we studied Jarlskog's invariant phases of unitary matrices 
instead of the determinant. 
Then we gave some useful formulas for them 
and derived the detailed dependency of them. 
Mathematically, a generalization of proposition \ref{prop:relation} 
is an interesting problem. It is a future task. 

 \section*{Acknowledgements}
The author is very grateful to Kazuyuki Fujii for helpful suggestion and 
comments on an earlier draft on this paper. 

\end{document}